\documentclass[a4paper,12pt]{article}
\usepackage{amssymb,amscd,amsmath}
\usepackage{graphicx}

\def\j#1#2#3#4{{\it #1} {\bf #2} #3 #4}
\def\be{\begin{equation}}
\def\ee{\end{equation}}
\def\bea{\begin{eqnarray}}
\def\eea{\end{eqnarray}}
\def\ie{\textit{i.e.} }

\title{Scientific and Financial Performance Measure : A Simultaneous Model to Evaluate Scientific Activities}

\author{L.T. Handoko\footnote{Email : handoko@lipi.fisika.net}}
\date{}

\begin{document}

\maketitle

\begin{picture}(0,0)
       \put(310,160){FISIKALIPI-05002}
\end{picture}

\thispagestyle{empty}

\begin{center}
\begin{small}
\noindent
Group for Theoretical and Computational Physics, Research Center for Physics, Indonesian Institute of Sciences\footnote{http://www.fisika.lipi.go.id}, Kompleks Puspiptek Serpong, Tangerang 15310, Indonesia \\
\end{small}
\end{center}

\vspace*{5mm}

\begin{abstract}
An alternative model to measure simultaneously scientific and financial performances of scientific activities is proposed. This mathematical model focuses only on the final scientific outcomes in each fiscal year to gurantee the objectivity. The model is suited for the purpose of immediate and quantitative evaluation needed by policy makers to make decision in the subsequent fiscal year. The model can be applied to any branches of science, while it is also adjustable to varying macro-economic indicators. This enables the policy makers to evaluate equally scientific activities in various fields of science. It is argued that implementing the model could realize a fair, transparent and objective reward and punishment system in any scientific activities in order to improve both individual and institutional performances. The model also enables an automatic evaluation embedded in any scientific databases either in the local system or over the net. 
\end{abstract}

\vspace*{5mm}

\newpage

\section{Introduction}
\label{sec:intro}

Management is an important aspect of human-being and its activities, including any kind of scientific activities. Management as the regulator and also the executor of regulation needs appropriate tools to implement the existing regulations. Regulation has been created as a common norm and commitment which should be followed by all (individual and institutional) fellows in order to establish a mutual relationship. Since mutual relationship is a crucial point to realize common purposes of all involved fellows and motivate them to work together. 

In the context of scientific activity, scientific management plays an important role, much more than another non-scientific (bureaucracy, business, etc) activities. This reflects the nature of scientific activity which is strongly based on the individual with unlimited degree of freedom. Because independency and freedom are necessary conditions for any scientific activities. On the other hand since a scientific activity is mostly supported by public fund, all scientific fellows and its activities should be accountable for public. In a real life, however fully transparent, easily-understandable and accountable scientific activities are often difficult to be realized. This is moreless due to the nature of scientific activity which is in most cases invisible and unpredictable. Fortunately, in contrast with non-scientific activities, scientific activities are always supposed to generate objective and measureable outcomes in a period of duration. 

This means there is an urgent niche on a specific tool to measure scientific performance based on the scientific outcomes. Because scientific outcome is the only element which is measureable. Moreover, performance measurement is not only limited to scientific performance, but also takes into account its economic potential. So far, scientific activities are almost justified by the aspect of economic potential using the method of cost-benefit analysis \cite{bozeman,kostoff}. While the scientific performance is measured naively based on the scientific outcomes. These methods clearly separate scientific and financial aspects, although both aspects are closely related each other. More than that, in daily practice it is hard to measure the objectivity on cost-benefit method since it adopts somehow absurd references as future potential which is yet unpredictable. 

It is also known more complete method as scientific and technical human capital (STHC) \cite{becker,schulz,polanyi,polanyi1,bozeman1,bourdieu,bourdieu1,coleman,coleman1} which observes a lot of aspects. Though the method looks ideal, this model is very complicated and involves a lot of subjective parameters. This reduces the accuracy and validity of the method as an easy, transparent and objective measurement tool. 

Therefore it is clear that as long as concerning scientific activity, performance should be measured based on the achievement of scientific outcome without considering its process. Inversely, this point makes scientific activity is easy to measure and quantize. Motivated by this fact, I propose an alternative tool in this paper called the Scientific and Financial Performance Measure (SFPM) Model.

The paper is organized as follow. In Sec. \ref{sec:teori} all assumptions introduced in the model and its theoretical aspects are given, followed by an example on how the model should be applied in the field of physical sciences in Sec. \ref{sec:aplikasi}. Before concluding the paper, a detail quantitative analysis is described in Sec. \ref{sec:analisa}.

\section{Theoretical framework}
\label{sec:teori}

Now let us first discuss the assumptions required in the model. As mentioned above, the first assumption in the model is all measurements are based only on scientific outcomes regardless its process. The scientific outcome itself is defined as : all outcomes generated in a scientific activity which have been recognized by independent third parties in a form of either scientific documents or other real activities. 

Secondly, the measurement is done in a year basis, \ie it takes into account all scientific outcomes in a fiscal year. This assumption is required since the tool is intended to measure the performances compared to the total budget granted in a fiscal year.

Next assumption is, each outcome is ranked based on its "difficulty to accomplish", and is then assigned with an appropriate scientific point ($S_P$) based on its "scientific weight" representing the "scientific importance". The order number ($N_O$) of all relevant scientific outcomes must be in order without duplication. On the other hand, the scientific point of an outcome may be the same with another neighboring scientific outcomes, but it must be smaller (greater) than another outcomes with different points above (below). This differentiates scientific outcomes in term of its financial contributions though the scientific importances might be comparable. Determination of the orders and the points of scientific outcomes may differ depend on the nature of each field of science. 

Further, it is also assumed some parameters as maximum scientific point ($P_M$), descending rate of scientific point ($P_D$, in percents) and total scientific point treshold per-scientist ($P_T$). These parameters should be the same for all fields of science. This method then enables a universal evaluation and comparation among different branches of science. It should be remarked that the absolute value of scientific point itself is less important, because it only represents the scale of discrepancies among scientific outcomes. Once the maximum scientific point and its descending rate have been determined, a series of available scientific points for each scientific outcome can be obtained through the formula, 
\be
         S_P = \left\{ P_M , P_D \times P_M , P_D^2 \times P_M , \cdots , P_D^{n_O-1} \times P_M , P_D^{n_O} \times P_M \right\} \; , 
        \label{eqn:sp}
\ee
where $n_O$ denotes the number of relevant scientific outcomes in a field. Remark that one should take the integer of $(S_P)_i$ for the sake of simplicity. Concerning the appropriateness, it is natural to restrict the scales of parameters for instance to be in order of, 
\be
        P_M  >  100 \; \; , \; \; 
        P_T  >  100 \; \; , \; \; 
        P_D  =  (50 \sim 90) \% \; .
\ee
if one puts the minimum value of $S_P$ should be greater than $1$ and $n_O > 10$. Namely if one takes $P_M = 10$ in this case, then the number of available $(S_P)_i$ is too small compared with $n_O$ and leads to a difficulty in assigning the scientific point for each relevant scientific outcome. This will be clarified through an illustration given in Sec. \ref{sec:aplikasi}. While the parameter $P_M$ works just as an overwhole scaling factor, it should be remarked that $P_T$ and $P_D$ are not absolute and subject to further detail analysis. However, a general prescription to determine these parameters quantitatively will be discussed in detail in Sec. \ref{sec:analisa}. 
 
Economic aspect related to financial performance is represented by the economic coefficient ($C_E$) which holds for all branches of science in a national scope. This parameter should be determined initially, and thereafter can be made varying automatically related to the macro economic indicators as inflation, economic growth, currency rate, etc. 

Finally, using all assumptions introduced above one can consider the ratios to measure the scientific and financial performances. For this purpose, I propose the following formulae,
\be
        R_S \equiv \frac{1}{n_P \times P_T} \sum_{i=1}^{n_O} 
        \left[ \left( S_P \right)_i \times \left( Q_O \right)_i \right] \; ,
        \label{eqn:rs}
\ee
to represent the ratio of scientific performance with $n_P$ denotes the number of scientists involved in the collaboration which generates the outcomes and $Q_O$ is the quantity of each scientific outcome. On the other hand, the ratio of financial performance is given by, 
\be
        R_F \equiv \frac{C_E}{B_T} \sum_{i=1}^{n_O} 
        \left[ \frac{\left( S_P \right)_i \times \left( Q_O \right)_i}{\left( N_O \right)_i} \right] \; ,
        \label{eqn:rf}
\ee
where $B_T$ is the total budget granted to the activity. From these equations, it is clear that $R_F$ is related directly to the scientific point and the order number of each scientific outcome. 

Of course the financial performance here measures only the "indirect" financial outcomes generated in a scientific activity or by a scientist in a fiscal year. However incorporating the "direct" financial outcomes, if any, is rather trivial, that is just adding it up to $R_F$. This means the total ratio for financial performance $R_F^T$ becomes, 
\be
        R_F^T \equiv R_F + \frac{F_O}{B_T} \; ,
\ee
where $F_O$ is the total amount of direct financial outcomes in the same unit as $B_T$ and $C_E$. 

As shown in Eq. (\ref{eqn:rs}), the model ignores the weight of each member in a scientific collaboration or institution which generates the scientific outcomes. This means, every member is awarded the same scientific point. This simplification is taken to avoid further subjectivity in the evaluation. However, this point might be relevant only when one evaluates the performance of individual scientist. 

Now we are ready to apply the model in a specific field of science to get the feeling on how the model should be applied and could be usefull. 

\section{Application} 
\label{sec:aplikasi}

From research management perspective, the model suggests two levels of regulators to keep the fairness and objectiveness in determining all parameters introduced above, that is
\begin{enumerate}
\item \underline{The official bureaucracy with an authority on scientific activity}.\\
The management in this level is responsible for determining the values of global parameters which hold for all branches of scientific activities, \ie $P_M$, $P_D$, $P_T$ and $C_E$. 
\item \underline{The relevant scientific community}. \\
The scientific community in a specific field is responsible for determining the order and assigning the appropriate values of scientific point according to its order for all relevant scientific outcomes in the field. Because these parameters are unique for a specific field. 
\end{enumerate}

Once all parameters are fixed, the system is ready for evaluating previous scientific activities and also decision-making of future scientific activities as well. Here, let us illustrate and apply the model in the field of physical sciences. Because the global physics community has already several comprehensive databases comprising a lot of scientific outcomes available on the net which are very usefull in this study. Taking simple statistics of the published papers in peer-reviewed journals, conference proceedings and so on, it has been found that concerning relevant scientific outcomes in the field as shown in the second column in Tab. \ref{tab:sp}, the appropriate scale for maximum scientific point should be $P_M = 200$. Again it does not matter if one takes 300 or even 400 for $P_M$, since it only reflects the whole scale of evaluation. The relevant scientific outcomes listed in the table are all kind of outcomes retrieved from three major databases in physics \cite{slac,arxiv,adsnasa}. Putting for instance the descending rate $P_D = 75\%$ and using Eq. (\ref{eqn:sp}), one obtains a series of available values for scientific points,  
\be
   S_P = \left\{ 200, 150, 112, 84, 63, 47, 35, 26, 19, 14, 10, 7, 5, 3, 2, 1 \right\} \; .  
\ee
Assuming the scientific points are taken as written in the third column and the economic coefficient has been namely taken to be US\$ 100, one can directly calculate the financial points ($F_O$) using the convertion formula for a single outcome, \ie $n_O = 1$, 
\be
        \left( F_O \right)_i = \frac{C_E \times (S_P)_i}{(N_O)_i} \; .
\ee
The results are  written in the last column for the order numbers and the scientific points assigned in the first and third columns in Tab. \ref{tab:sp}. 

Utilizing Tab. \ref{tab:sp} and Eqs. (\ref{eqn:rs}) and (\ref{eqn:rf}), one can immediately calculate the whole performance of an individual scientist, a single scientific project and a scientific institution using their real outcomes in a fiscal year. Since everything is simply mathematics, once the global parameters and the order of scientific outcomes for a specific field have been determined, the rest can be done automatically to get the total score. For example if a physicist, granted with a totally US\$ 5,000 research fund, has published two papers in major international journals and one book through an international publisher, then the scientific and financial performances in that fiscal year become, 
\be
        \begin{array}{lcll}
        R_S & = & \displaystyle \frac{(150 \times 1) + (112 \times 2)}{1 \times 250} & \times 100 \% = 109 \% \; , \\
        R_F & = & \displaystyle \frac{(3,750 \times 1) + (1,600 \times 2)}{5,000} & \times 100 \% = 139 \% \; , 
        \end{array}
\ee
for $P_T = 250$. In this case only the indirect financial outcomes have been taken into account since the direct one is absent. The same procedure should be taken to obtain the performances for an institution or a single project.

\begin{table}[t]
        \begin{center}
        \begin{tabular}{||r|p{9cm}|c|rr||}
        \hline\hline
        \multicolumn{1}{||c|}{NO} & \multicolumn{1}{|c|}{SCIENTIFIC OUTCOME} & \multicolumn{1}{|c|}{$S_P$} & \multicolumn{2}{|c||}{FINANCIAL} \\
        \hline
        1. & Foreign patent & 200 & US\$ & 20,000,- \\
        2. & International scientific award & 200 & US\$ & 10,000,- \\
        3. & Local patent & 150 & US\$ & 5,000,- \\
        4. & Book published by foreign publisher & 150 & US\$ & 3,750,- \\
        5. & National scientific award & 150 & US\$ & 3,000,- \\
        6. & Copyright & 112 & US\$ & 1,867,- \\
        7. & International regular journal & 112 & US\$ & 1,600,- \\
        8. & Trademark & 84 & US\$ & 1,050,- \\
        9. & Supervising passed PhD disertation & 63 & US\$ & 700,- \\
        10. & Book published by local publisher & 63 & US\$ & 630,- \\
        11. & International proceeding & 47 & US\$ & 427,- \\
        12. & Pupular article in foreign media & 47 & US\$ & 392,- \\
        13. & Book translation & 47 & US\$ & 362,- \\
        14. & Supervising passed MSc theses & 35 & US\$ & 250,- \\
        15. & Invited speaker & 26 & US\$ & 173,- \\
        16. & Trainer & 26 & US\$ & 163,- \\
        17. & Supervising passed BSc theses & 19 & US\$ & 112,- \\
        18. & Local regular journal & 14 & US\$ & 78,- \\
        19. & Local proceeding & 10 & US\$ & 53,- \\
        20. & Popular article in local media & 7 & US\$ & 35,- \\
        \hline\hline
        \end{tabular}\\
        \caption{An example of relevant scientific outcomes in the field of physics science, the scientific points and the financial convertions for $P_M = 200$, $P_D = 75\%$ and $C_E =$ US\$ 100,-. }
        \label{tab:sp}
        \end{center}
\end{table}

For the order of scientific outcomes and its scientific points one can rely on the common sense recognized by relevant scientific community in the field. So the remaining problem is how to reduce the subjectivity in determining the global parameters. This problem stays on the official bureaucracy as mentioned above who might not be familiar with the scientific standards. In the next section a general prescription to deal with this problem is discussed. 

\section{Further analysis}
\label{sec:analisa}

\begin{figure}[t]
        \begin{minipage}{75mm}
        \centering \includegraphics[width=7cm]{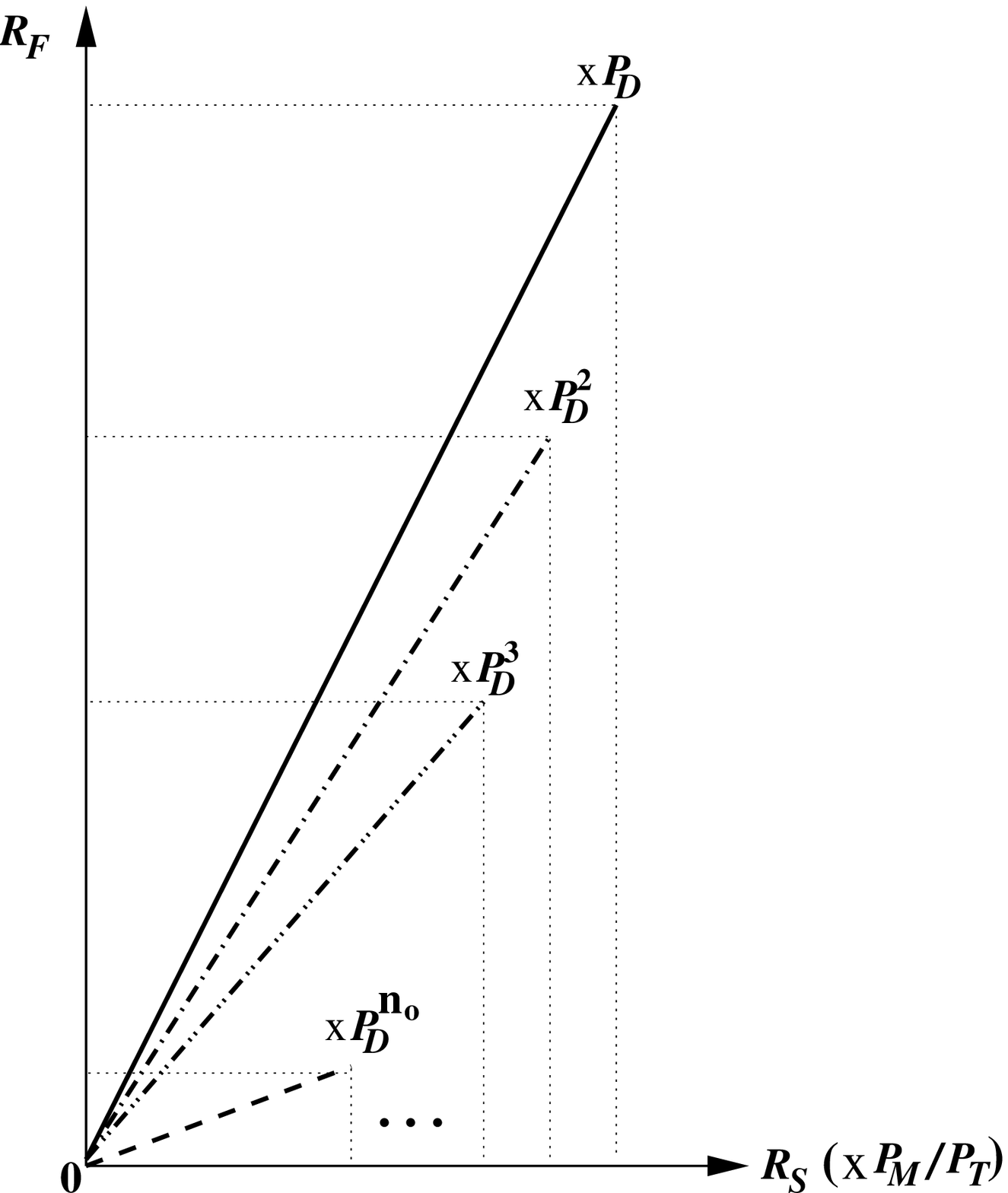}
        \caption{Correlation between $R_F$ and $R_S$ for each scientific outcome per-person. }
        \label{fig:korelasi}
        \end{minipage}
        \hspace*{3mm}
        \begin{minipage}{75mm}
        \centering \includegraphics[width=6cm]{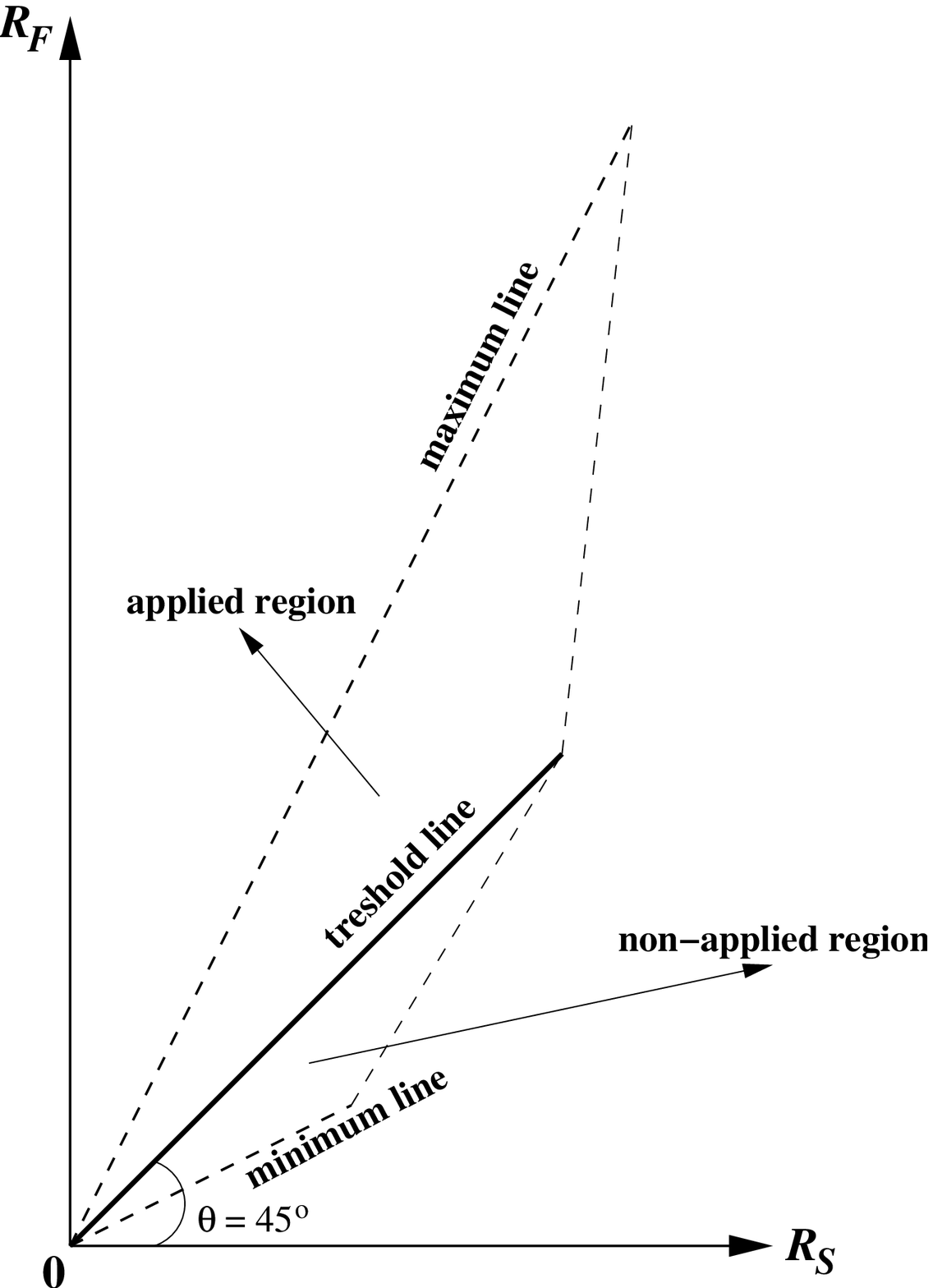}
        \caption{Treshold between the appliend and non-applied sciences in term of its scientific and financial performances. }
        \label{fig:treshold}
        \end{minipage}
\end{figure}

At this stage, one can obtain easily the correlation between the scientific and financial performances using Eqs. (\ref{eqn:rs}) and (\ref{eqn:rf}) as follows, 
\be
        \left( R_F \right)_i = \tan \theta_i \times \left( R_S \right)_i \; ,
\ee
for each outcome. The slope is determined by the angle $\theta_i$ which satisfies the relation,
\be
        \tan \theta_i = \frac{C_E \times P_T \times n_P}{B_T} \frac{1}{\left( N_O \right)_i} \; .
        \label{eqn:angle}
\ee
This correlation function is depicted in Fig. \ref{fig:korelasi} for the 1st till the $n_O-$th scientific outcome with $n_P = 1$. The horizontal length for each outcome is determined by $\left( R_S \right)_i =\left( P_D \right)^i \times \left( {P_M}/{P_T} \right)$. 

The allowed region in the $R_S-R_F$ plane is given in Fig. \ref{fig:treshold}. One can roughly divide the region into two areas, the upper and lower areas from the treshold line. The treshold line shows the equilibrium condition where the scientific and financial performances are completely equal, \ie $\tan \theta_\mathrm{treshold} = 1$. It can be argued that this line might be used to determine whether a scientific outcome relative to its total budget is categorized as applied (financial oriented) or non-applied (scientific oriented) outcome.  This mathematical definition of applied and non-applied outcome is new and provides a new way to categorize the outcomes in any scientific activities. 

From this point of view, the definition of "applied" and "non-applied" are rather mathematical than subject to wide interpretation as the conventional view. This means, as experienced in daily life, the interpretation of applied and non-applied outcomes are non-trivial and dynamic depending on how much the total budget ($B_T$) and how many person are needed to generate the outcomes. This result generalizes common sense of "applicability", but it is actually more natural. Since something is "applicable" if someone is able to accomplish it with less budget than another ones which require more regardless how much the potential revenue can be obtained in the future which is always the subject of subjectivity.

Moreover as shown in Eq. (\ref{eqn:angle}), this result also proves quantitatively that the number of scientists in a collaboration is proportional to the level of "applicability" of their outcomes. Actually one can observe that, for instance theoretical studies which are less applicable are generally worked out by groups with less people than the experimental ones. 

On the other hand, the treshold line could be utilized to determine the values of $P_D$. One first puts a specific scientific outcome, which can be clearly categorized as either applied or non-applied outcome for a reference value of $B_T$,  to be on the treshold line. Thereafter, another scientific outcomes can be put into one of two regions relative to the previous one. The parameter $P_D$ then can inversely be extracted such that all outcomes stay on the desired regions for given values of $P_M$ and $P_T$. $P_T$ can be obtained in advanced through statistic analysis on some existing databases of scientific outcomes in the related field. This prescription at least would significantly reduce the ambiguity in determining the global parameters as mentioned in Sec. \ref{sec:teori}.

\section{Conclusion and discussion}
\label{sec:diskusi}

A new alternative tool, the SFPM Model,  to evaluate scientific and financial performances has been discussed. In the model, a simultaneous relation between scientific and financial performances has been formulated assuming that the quantitative scores should be extracted only from the scientific outcomes. The model solves a crucial problem on how to measure simultaneously both scientific and financial performances of scientific activities in various fields. It provides a simple tool for immediate evaluation which is in practical daily management urgently needed. Since the model is based on the completely quantitative measurements, it could avoid any ambiguities and then guarantees the objectiveness of evaluation process. A sustainable evaluation utilizing the model could also measure and integrate long-term scientific and financial performances in complement with the other known evaluation methods. 

Here, I list several advantages of measurement tool based on the SFPM Model, 
\begin{itemize}
\item Since it is based only on the scientific outcomes, the objectivity and transparency of measurement can be guaranteed. 
\item A single year base method makes the evaluation process and decision-making for next fiscal year easier, since the result reflects an up-to-date real condition. 
\item In long term, the whole annual evaluations of each fellow can be compiled to implement better compensation system. For example, the evaluation result can be used as a basic reference to distribute and allocate human resources and funding to each individual or institution. 
\item The same tool can be used to evaluate the advisability of incoming proposals in term of the promising targets claimed in. 
\end{itemize}

Implementing the SFPM measurement tool consistently and continously in long term for each individual, scientific project and institution could generate performance indicators in each level. This has a great potential to help identifying the fundamental problems related to scientific activities in a regional or national scale. Combining the result with a consistent compensation system could also improve the whole scientific performance. 

Most of the parameters introduced above are not absolute and only indicate the scale, which make them easy to be determined without requiring deep consideration. However, more serious consideration is still required for few of them, which especially constitute all fields absolutely as descending rate of scientific point ($P_D$) and total scientific point treshold per-scientist ($P_T$). The parameter $P_T$ can be determined in a comprehensive way by for instance taking the average of all scientific outcomes generated by scientists around the world. The data can be retrieved easily through the existing databases available on the net. This result further can be used together with the treshold line method described in Sec. \ref{sec:analisa} to determine quantitatively the appropriate descending rate $P_D$. Therefore, further research utilizing available databases providing global \cite{slac,arxiv,adsnasa,google} and local \cite{dbriptek} scientific outcomes on the net is highly recommended. 

Finally, we would like to notice that an online calculator applying this model is now under construction. Technically, this kind of online tool can be embedded in any existing databases (patent, bibliography, etc) to enable a full automatic evaluation process for research institutions and scientists around the world. An example with the scope of Indonesian scientific community is still under progress through the DBRIpTek database \cite{dbriptek,ocsp}.

\section*{Acknowledgment}

We would like to appreciate D. Sajuti for providing the problem which initially motivated this work, and A. Kusnowo for fruitfull discussion during the final stage of this work. This project is partly funded by Riset Kompetitif LIPI (fiscal year 2005).


\begin{thebibliography}{99}
        \bibitem{bozeman} B. Bozeman and J. Melkers, 
        \j{Evaluating R\&D impacts: Methods and practice, Boston, Kluwer Academic Publishers}{}{(1993).}{}.
        \bibitem{kostoff} R. Kostoff, H. Averch and D. Chubin, 
        \j{Evaluation Review}{18}{(1994)}{3}. 
        \bibitem{becker} G. Becker, 
        \j{Journal of Political Economics}{70}{(1962)}{S9}.
        \bibitem{schulz} T. W. Schultz, 
        \j{The economic value of education, New York, Columbia University Press}{}{(1963).}{}.
        \bibitem {polanyi} M. Polanyi, 
        \j{The tacit dimension, London, Cox \& Wyman}{}{(1967).}{}.
        \bibitem{polanyi1} M. Polanyi, 
        \j{The logic of tacit inference : Knowing and being, London, Routledge \& Kegan Paul}{}{(1969).}{}.
        \bibitem{bozeman1} B. Bozeman, 
        \j{Policy Studies Journal}{14}{(1986)}{519}.
        \bibitem{bourdieu} P. Bourdieu, 
        \j{Handbook of theory and research for the sociology of education New York, Greenwood}{}{(1986).}{}.
        \bibitem{bourdieu1} P. Bourdieu and L. Wacquant, 
        \j{An invitation to reflexive sociology, Chicago, University of Chicago Press}{}{(1992).}{}
        \bibitem{coleman} J. C. Coleman, 
        \j{American Journal of Sociology}{94}{(1988)}{S95}.
        \bibitem{coleman1} J. C. Coleman, 
        \j{Foundations of social theory, Cambridge, The Belknap Press of Harvard University Press}{}{(1990).}{} 
         \bibitem{slac} SLAC Spires, 
                \j{http://www.slac.stanford.edu/spires/.}{}{}{}
        \bibitem{arxiv} arXiv.org - e-Print archive, 
                \j{http://www.arxiv.org.}{}{}{}
        \bibitem{adsnasa} Astrophysics Data System - NASA ADS,
                \j{http://adsabs.harvard.edu.}{}{}{}
        \bibitem{google} Google Scholar, 
                \j{http://scholar.google.com.}{}{}{}
        \bibitem{dbriptek} DBRIpTek - Database Riset, Ilmu Pengetahuan dan Teknologi, 
                \j{http://www.dbriptek.lipi.go.id}{}{(2003).}{}
        \bibitem{ocsp} OCSP - Online Calculator for Scientific Performance, 
        \j{http://www.koki.lipi.go.id}{}{(2005).} 
\end{thebibliography}
\end{document}